\documentclass[pra,aps,twocolumn]{revtex4}
\usepackage{amssymb}
\usepackage{latexsym}
\usepackage{graphicx}
\usepackage{boldtensors}
\newcommand{\Z}{{\mathbb Z}}

\newcommand{\R}{{\mathbb R}}

\newcommand{\LL}{{\mathbb L}}

\def\be{\begin{equation}}
\def\ee{\end{equation}}
\def\bea{\begin{eqnarray}}
\def\eea{\end{eqnarray}}

\def\d{{\,\rm d}}

\def\bfz{{\bf 0}}

\def\k{{\bf k}}
\def\e{{\bf e}}
\def\g{{\bf g}}

\def\q{{\bf q}}

\def\p{{\bf p}}

\def\v{{\bf v}}
\def\x{{\bf x}}

\def\y{{\bf y}}

\def\K{{\bf K}}
\def\P{{\bf P}}

\def\Q{{\bf Q}}
\def\X{{\bf X}}

\def\h2m{\frac{\hbar^2}{2m}}

\begin{document}
\title{
\vspace{1cm} \large\bf Total momentum and thermodynamic phases of quantum systems\\}
\author{Andr\'as S\"ut\H o\\
Institute for Solid State Physics and Optics, Wigner Research Centre for Physics,
Hungarian Academy of Sciences, P. O. Box 49, H-1525 Budapest,  Hungary}
\thispagestyle{empty}
\begin{abstract}
\noindent
The total momentum of $N$ interacting bosons or fermions in a cube equipped with periodic boundary conditions is a conserved quantity. Its eigenvalues follow a probability distribution, determined by the thermal equilibrium state. While in non-interacting systems the distribution is normal with variance $\sim N$, interaction couples the single-particle momenta, so that the distribution of their sum is unpredictable, except for some implications of Galilean invariance. First, we present these implications which are strong in 1D, moderately strong in 2D, and weak in 3D. Then, we speculate about the possible form of the distribution in fluids, crystals, and superfluids. The existence of phonons suggests that the total momentum can remain finite when $N\to\infty$. We argue that in fluids the finite momenta distribute continuously, but their integrated probability is smaller than 1, because the momentum can also tend to infinity with $N$. In the fluid-crystal transition we expect that the total momentum becomes finite with full probability and distributed over a lattice, and that in the fluid-superfluid transition a delta peak appears only at zero total momentum. Based on this picture, we discuss the superfluid flow in both the frictionless and the dissipative cases, and derive a temperature-dependent critical velocity. Finally, we show that Landau's criterion for excitations in moving superfluids is an in some cases correct result of an erroneous derivation.

\vspace{2mm}\noindent PACS: 03.75.Kk, 05.30.-d, 67.10.Fj, 67.25.dj, 67.80.bd

\end{abstract}
\maketitle
Our concern in this paper is to understand the role of the total momentum in the description of fluids, crystals and superfluids. The idea that such a role can exist at all is based on the experimental fact that in interacting quantum systems there are excited states that involve a large number of particles and carry a finite total momentum. Such collective modes exist at all energies, in all the thermodynamic phases; we think of compression waves or lattice vibrations~\cite{Fey1}. This fact is in a striking contrast with the naive expectation that, because the total momentum is the sum of $N$ single-particle momenta, it should be normally distributed with a variance $\sim N$. Deviation from this can be a joint effect of interaction and quantum mechanics. In interacting quantum systems the single-particle momenta are not independent: because they do not commute with the interaction, free energy minimization couples them. Therefore, the distribution of their sum is largely unpredictable, although we shall see that Galilean invariance does have implications for it. In classical systems the single-particle momenta are identically distributed independent random variables, so their sum is normally distributed with variance $\sim N$, and similar holds in non-interacting quantum systems.

Once the total momentum can be either finite or tending to infinity with $N$, one can speculate about the changes that can take place in phase transitions. Two characteristic changes are to be expected when the temperature $T$ decreases. First, the continuous distribution for the finite momentum, certainly existing at high $T$, can partly or entirely be replaced by a discrete one. Second, the total momentum of order $\sqrt{N}$, also present at high $T$, can disappear. We shall argue that both occur in the fluid-solid transition, while in the fluid-superfluid transition only a delta peak appears at zero total momentum. We do not expect a qualitative change in the vapor-liquid transition.

A crucial point that must be stressed is that the total momentum is a property of the entire system. Saying that in He II below the $\lambda$-point it can be zero with probability $p$ and nonzero with probability $1-p$ will mean that the equilibrium state is a convex combination of a superfluid and a normal fluid state with respective weights $p$ and $1-p$. This is the least intuitive aspect of our proposal: that a physical meaning is attributed not to parts of the system (e.g. to $N_0<N$ particles forming the Bose condensate) but to parts of the density matrix. A detailed account of this work is given elsewhere~\cite{Su1}.

\vspace{3pt}
\noindent
{\em Total momentum and Galilean invariance.}--As in our earlier paper~\cite{Su2}, we consider the energy operator
\be
H=\frac{1}{2m}\sum_{j=1}^N \p_j^2+U_\Lambda(\x_1,\ldots,\x_N)
\ee
of $N$ interacting particles in a $d$-dimensional cube $\Lambda$ of side length $L$, defined with periodic boundary conditions, which admit Galilean transformation in a bounded domain. $U_\Lambda$ is the periodized potential energy containing translation invariant pair and possibly many-body interactions, 
which can include hard-core interactions but no external field. In such a case the total momentum $\P=\sum_j\p_j\equiv -i\hbar\sum_j \partial/\partial\x_j$ is conserved and the eigenvalues and eigenvectors of $H$ can be labeled by the eigenvalues $\hbar\Q$ of $\P$: $\P\psi_{\Q,n}=\hbar\Q\psi_{\Q,n}$, $H\psi_{\Q,n}=E_{\Q,n}\psi_{\Q,n}$,
where $E_{\Q,n}\leq E_{\Q,n+1}\ (n\geq 0)$ and
\be
\Q\in\Lambda^*=\{(2n_1\pi/L)\e_1+\cdots+(2n_d\pi/L)\e_d\}_{n_1,\ldots,n_d\in\Z}.
\ee
$\psi_{\bfz,0}$ and $E_{\bfz,0}$ are the ground state and its energy. A $N^d$-point subset of $\Lambda^*$, that we call irreducible,
\be\label{lambdairred}
\Lambda^*_{\rm irred}=\{\q\in\Lambda^*: -\pi N/L<q_i\leq\pi N/L,\ i=1,\ldots,d\},
\ee
plays a distinguished role: any $\Q\in\Lambda^*$ can uniquely be written as $\q+N\k$, where $\q\in\Lambda^*_{\rm irred}$ and $\k\in\Lambda^*$, and all the eigenvalues and eigenvectors are obtained by a Galilean boost from those belonging to $\Lambda^*_{\rm irred}$~\cite{Su2},
\bea\label{Galilei}
E_{\q+N\k,n}&=&E_{\q,n}+(\hbar^2/2m)[N\k^2+2\k\cdot\q],\nonumber\\
\psi_{\q+N\k,n}&=&e^{iN\k\cdot\overline{\x}}\psi_{\q,n},\quad \q,\k\in\Lambda^*, n\geq 0,
\eea
where $\overline{\x}=N^{-1}\sum_{j=1}^N\x_j$, the center of mass. With $\beta=(k_BT)^{-1}$ and $Z=\sum_\Q\sum_n e^{-\beta E_{\Q,n}}$, the density matrix is $Z^{-1}\sum_\Q\sum_n e^{-\beta E_{\Q,n}}|\psi_{\Q,n}\rangle\langle\psi_{\Q,n}|$. The sum
\be
\nu_\Q=Z^{-1}\sum_{n=0}^\infty e^{-\beta E_{\Q,n}}
\ee
is interpreted as the probability that $\P=\hbar\Q$ in thermal equilibrium. Taking the thermodynamic limit at a fixed density $\rho=N/L^d$, one can prove the following~\cite{Su1}.

\noindent
(i) In 1D, with $\lambda_\beta=\sqrt{2\pi\beta\hbar^2/m}$,
\be\label{CLT1D}
\lim_{N\to\infty}{\rm Prob}\left\{\frac{|Q|}{\rho\sqrt{N}}\leq x\right\}
=\frac{\lambda_\beta\rho}{\pi}\int_0^x e^{-\frac{\lambda_\beta^2\rho^2}{4\pi}y^2}\d y.
\ee

\noindent
(ii) In 2D, for any $J>0$ integer,
\be\label{low-2D-up}
\lim_{N\to\infty} {\rm Prob}\left\{\frac{|Q_i|}{\sqrt{\rho N}}>(2J-1)\pi\right\}\sim e^{-{\rm const.}\times J^2}.
\ee

\noindent
(iii) For $d\geq 3$,
\be\label{geq-3}
\lim_{N\to\infty}{\rm Prob}\left\{|Q_i|\leq 3\pi N/L\right\}=1
\ee
[but we believe that $|\Q|/\sqrt{N}\to\infty$ has a vanishing asymptotic probability, just as for $d=1,2$].

One can add that for non-interacting systems a normal distribution of variance $\sim N-N_\bfz$ [$N_\bfz$ is the number of particles with zero momentum] is expected in any dimension. The specificity of 1D is that the irreducible wave numbers are in the interval $(-\pi\rho,\pi\rho]$ independently of the system size. Because the interaction fully reveals itself in the irreducible ensemble, this gives a hint to why, on the scale $Q\sim\sqrt{N}$, do any 1D system resemble a non-interacting one~\cite{rem4}. On the other hand, $\max_{\Q\in\Lambda^*_{\rm irred}}|Q_i|\propto\sqrt{N}$ in 2D and $\propto N^{2/3}$ in 3D; for this reason one cannot expect that the analog of Eq.~(\ref{CLT1D}) would hold for interacting systems in 2D and 3D.

\vspace{3pt}
\noindent
{\em Density matrix reduced to the center of mass.}--$\P$ is canonically conjugate to $\overline{\x}$. A consequence is that the dependence on $\overline{\x}$ can be separated in $\psi_{\Q,n}$: with the notations $\X=(\x_1,\ldots,\x_N)$, $\X'=(\x'_2,\ldots,\x'_N)$, $\x'_j=\x_j-\x_1$, 
\be\label{psix-separated}
\psi_{\Q,n}(\X)=e^{i\Q\cdot\overline{\x}} e^{-(i/N)\Q\cdot\sum_{j=2}^N\x'_j}\psi_{\Q,n}(\bfz,\X').
\ee
We can reduce the density matrix to the center of mass through a partial trace over $\X'$~\cite{Het}, with the result
\be\label{rhocm-kernel}
\langle\overline{\x}|\rho_{\rm c.m.}|\overline{\y}\rangle
=\rho\sum_{\Q\in\Lambda^*}\nu_\Q e^{i\Q\cdot(\overline{\x}-\overline{\y})},\
\langle\Q|\rho_{\rm c.m.}|\Q'\rangle=N\delta_{\Q,\Q'}\nu_\Q,
\ee
so $N\nu_\Q$ are the eigenvalues of $\rho_{\rm c.m.}$, with corresponding eigenfunctions $\langle\overline{\x}|\Q\rangle=L^{-\frac{d}{2}}e^{i\Q\cdot\overline{\x}}$.

Equations~(\ref{Galilei})-(\ref{rhocm-kernel}) are valid at all $T$ and $\rho$, for bosons and fermions, with or without interaction. From now on, we deal only with bosons. We recall that the one-particle reduced density matrix is
\be\label{rho1-kernel}
\langle\x|\rho_1|\y\rangle=\rho\sum_{\Q\in\Lambda^*}n_\Q e^{i\Q\cdot(\x-\y)},\ \langle\Q|\rho_1|\Q'\rangle=N\delta_{\Q,\Q'}n_\Q
\ee
where $n_\Q=\langle N_\Q\rangle/N$ and $N_\Q$ is the occupation number operator for the single-particle state $|\Q\rangle$. Because $\sum_{\Q\in\Lambda^*}n_\Q=1$ just as $\sum_{\Q\in\Lambda^*}\nu_\Q=1$, $n_\Q$ can be interpreted as the probability that a particle is in the one-particle state $|\Q\rangle$. Thus, there is a perfect structural analogy between $\rho_{\rm c.m.}$ and $\rho_1$. They have the same eigenvectors $|\Q\rangle$, the corresponding eigenvalues are probabilities multiplied by $N$, the expressions~(\ref{rhocm-kernel}) and (\ref{rho1-kernel}) are nonnegative, therefore $\nu_\Q$ and $n_\Q$ are positive definite functions on $\Lambda^*$: the $n\times n$ matrices $[\nu_{\Q_i-\Q_j}]_{ij}$ and $[n_{\Q_i-\Q_j}]_{ij}$ are positive semidefinite for any $n$ and any $\Q_1,\ldots,\Q_n$ in $\Lambda^*$~\cite{Su1}. In particular,
\be\label{qmax=0}
\nu_\bfz\geq\nu_\Q,\quad n_\bfz\geq n_\Q\quad\mbox{any $\Q\in\Lambda^*$}.
\ee
We note that $\nu_\bfz\geq\nu_\Q$ implies that the free energy density can be computed in the $\Q=\bfz$ ensemble~\cite{Su1}.

\vspace{3pt}
\noindent
{\em Distribution of the total momentum in infinite space.}--Structural similarity between $\rho_{\rm c.m.}$ and $\rho_1$ is contrasted with a fundamental difference, owing to the fact that the dependence of the eigenstates $\psi_{\Q,n}$ on $\overline{\x}$ is separable, but the dependence on $\x_j$ is not. The difference is accentuated in the thermodynamic limit in which both $\nu_\Q$ and $n_\Q$ tend in distribution sense to a positive and positive definite measure on the dual space $(\R^d)^*$. We denote them by $\tilde{\nu}(\k)$ and $\tilde{n}(\k)$, respectively. If there is Bose-Einstein condensation (BEC) then $\tilde{n}(\k)=\tilde{n}_\bfz\delta(\k)+\tilde{n}_c(\k)$ where $\tilde{n}_\bfz>0$ and $\tilde{n}_c$ is continuous. Now
$
\int\tilde{n}(\k)\d\k=1
$
always holds true, otherwise the kinetic energy density would diverge in the thermodynamic limit~\cite{Su3}. So $\tilde{n}$ is a probability measure, meaning that in infinite space, as in finite volumes, the single-particle momentum cannot be infinite. This is not true for $\tilde{\nu}$: in finite volumes the total momentum is finite with probability 1, but from Eqs.~(\ref{CLT1D}) and (\ref{low-2D-up}) one can infer that $p_{<\infty}\equiv\int\tilde{\nu}(\k)\d\k=0$ in 1D and $p_{<\infty}<1$ in 2D for all $T>0$; also, $p_{<\infty}<1$ should hold for high $T$ or small $\rho$ in 3D. $p_{<\infty}$ and $p_\infty\equiv 1-p_{<\infty}$ are the respective probabilities that the total momentum is finite or infinite in infinite volume.

$p_{<\infty}=0$ means that states carrying a finite total momentum, compression waves included, gradually lose all their statistical weight with the increasing system size. This occurs at $T>0$ in non-interacting systems in any dimension, and also in interacting systems in 1D. This does not mean that density waves as finite-energy excitations do not exist. However, in any dimension the states with an energy gap of order 1 above the ground state can play a role only at $T=0$: at $T>0$, $E_{\Q,n}-E_{\bfz,0}\propto N$ for the relevant eigenstates.

In contradistinction to collective excitations, few-particle excitations involve small groups of particles which are nearly separated from each other. Rotons may be of this kind~\cite{Fey}. In 3D at high energies, in 2D at all energies they are expected to occur with a positive density in a non-vanishing fraction of eigenstates, and their finite momenta to add up randomly to a total momentum of order $\sqrt{N}$.

\vspace{3pt}
\noindent
{\em Total momentum and thermodynamic phases: Speculations.}--Now we can make our proposals about the connection between $\tilde{\nu}$ and the thermodynamic phases. The above two classes of excitations, the collective ones with a finite nonzero total momentum and the single- and few-particle excitations with an accumulated infinite total momentum in infinite space are sufficient to characterize the fluid phases (gases and liquids) at $T>0$ and $d>1$. Namely, $\tilde{\nu}$ is continuous, and the equilibrium state ${\cal S}$ in infinite volume is of the form ${\cal S}=p_{<\infty}\ {\cal S}_{<\infty}+p_{\infty}\ {\cal S}_{\infty}$. Here ${\cal S}_{<\infty}$ and ${\cal S}_{\infty}$ themselves are composed of states with finite or infinite momentum, respectively, and can be obtained via suitable limits from parts of the density matrix~\cite{Su1}. Since in 2D the second term is nonzero at all $T>0$, a 2D system cannot be in a crystalline state (see below). This corresponds to known results about the preservation of shift-invariance~\cite{Mer,FP1,FP2,Rich}.

Concerning crystals, physical intuition suggests that the random single- and few-particle motions freeze out in the thermodynamic limit, implying $p_{<\infty}=1$. Moreover, the limit of $\langle\overline{\x}|\rho_{\rm c.m.}|\overline{\y}\rangle$ must be periodic in $\overline{\y}-\overline{\x}$ according to some lattice $\LL$, therefore $\tilde{\nu}(\k)=\sum_{\K\in\LL^*}\tilde{\nu}_\K\delta(\k-\K)$, where $\LL^*$ is the reciprocal lattice. Because of positive definiteness, $\tilde{\nu}_\bfz\geq\tilde{\nu}_\K$. Coherent quantum crystals~\cite{KN,N} are curious objects interpolating between crystals and fluids. For a discussion see Refs.~\cite{Su1}, \cite{Su3}.

The characterization of superfluids raises two questions: what is the role of BEC, and what are the two fluids in Tisza's two-fluid picture~\cite{Tisza}. There is today a consensus based on analytical, numerical and experimental results~\cite{Lon,PO,Cep,AZ,SS,Gly,Mor} that in bulk liquid helium BEC and the superfluid transition occur simultaneously. Yet in 2D there is superfluidity~\cite{Dal,KK} but no BEC at $T>0$~\cite{Hoh,Wag,BM}, and bulk helium at $T=0$ is 100\% superfluid containing less than 10\% of condensate. These controversial facts can be reconciled by assuming that both in 2D and in 3D the primary event at the superfluid transition temperature $T_s$ is 'condensation' of the entire system into a state of zero total momentum, and in 3D this triggers settling of individual particles at zero momentum. Thus, when passing  $T_s$, the continuous distribution $\tilde{\nu}$ picks up a Dirac-delta, so that it becomes $\tilde{\nu}(\k)=\tilde{\nu}_\bfz\delta(\k)+\tilde{\nu}_c(\k)$, where $\tilde{\nu}_\bfz=\lim_{\kappa\to 0}\lim_{L\to\infty}\sum_{|\q|\leq\kappa}\nu_\q>0$ and $\tilde{\nu}_c$ is continuous. An immediate implication is that the super and normal fluids are not parts (as the condensate) but alternative states of the whole system, that compose the equilibrium state by convex combination,
\be\label{s-n-decomp}
{\cal S}=\tilde{\nu}_\bfz(T)\ {\cal S}_s(T)+[1-\tilde{\nu}_\bfz(T)]\ {\cal S}_n(T).
\ee
The super and normal fluid fractions are $\tilde{\nu}_\bfz$ and $1-\tilde{\nu}_\bfz$, respectively. ${\cal S}_s$ is a state of zero total momentum, and ${\cal S}_n$ can further be decomposed according to whether the total momentum is nonzero finite, or infinite. For non-interacting bosons $\tilde{\nu}(\k)\equiv 0$ at $T>0$, so in spite of BEC, their state is not a superfluid.

\vspace{3pt}
\noindent
{\em Frictionless flow.}--Galilean boost generates frictionless flow. If the flow has a velocity $\v$ with respect to the container $\Lambda$, and $m\v/\hbar\in\Lambda^*$,
then in the comoving reference frame we have~\cite{Su1,Su2}
\bea\label{Galilei+}
H^\v\psi_{\Q,n} &=& E_{\Q-Nm\v/\hbar,n}\psi_{\Q,n},\nonumber\\
\P^\v\psi_{\Q,n} &=& (\hbar\Q-Nm\v)\psi_{\Q,n}.
\eea
$H^\v$ and $\P^\v$ are obtained from $H$ and $\P$, respectively, by replacing $\p_j$ with $\p_j-m\v$~\cite{rem3}. $H^\v$ has the same eigenvalues and eigenfunctions as $H$: the velocity boost only cyclicly permutes the eigenstates among the energies, by adding $Nm\v$ to their momentum. The density matrix reduced to the center of mass will be
$
\langle\bfz|\rho^\v_{\rm c.m.}|\x\rangle =e^{\frac{-iNm}{\hbar}\v\cdot\x}\langle\bfz|\rho_{\rm c.m.}|\x\rangle,
$
and its square root can play the role of a macroscopic wave function. For large $L$ and large $|\x|$, $\langle\bfz|\rho_{\rm c.m.}|\x\rangle\approx\rho\tilde{\nu}_\bfz$. The super and normal fluid states in infinite volume can be constructed, and Eq.~(\ref{s-n-decomp}) becomes
\be
{\cal S}^\v(T)=\tilde{\nu}_\bfz(T)\ {\cal S}^\v_s(T)+[1-\tilde{\nu}_\bfz(T)]\ {\cal S}^\v_n(T),
\ee
with unchanged super and normal fluid fractions. This is an equilibrium state whose free energy density is the same as that of the fluid at rest.

\vspace{3pt}
\noindent
{\em Dissipative flow: Speculations.}--A real flow is frictionless in the superfluid state but is dissipative in the normal fluid state.
If in the superfluid state the velocity is $\v$, then in the normal fluid state it will be smaller and time dependent, $\alpha(t)\v$, $\alpha(t)<1$. The difference is due to the produced heat which partly is dissipated into the environment, partly raises the temperature to $T_t>T$ in both states~\cite{rem}. Now ${\cal S}^\v(T)$ is replaced by
\be
{\cal S}^{\v}_{\rm noneq}(T_t)=\tilde{\nu}_\bfz(T_t)\ {\cal S}_{s}^\v(T_t)+\left[1-\tilde{\nu}_\bfz(T_t)\right]{\cal S}_{n}^{\alpha(t)\v}(T_t).
\ee
This is a non-equilibrium state because the two terms are generated by different Hamiltonians, $H^\v$ and $H^{\alpha(t)\v}$, respectively~\cite{Wre}. $T_t$ can be computed from
\be\label{TtT}
\frac{1}{2}k_BT_t-\frac{1}{2}k_BT=\eta \frac{mv^2}{2}[1-\alpha(t)^2],
\ee
where $\eta\leq 1$ is the efficiency, i.e., the fraction of the heat that raises the temperature. The dissipation continues until $\alpha(t)=0$. It raises $T_t$ and drives $\alpha(t)$ to zero. In the limit $t\to\infty$ the system goes over into a non-equilibrium steady state at a temperature $T_v$,
\be
{\cal S}^{\v}_{\rm noneq}(T_v)=\tilde{\nu}_\bfz(T_v)\ {\cal S}_{s}^\v(T_v)+\left[1-\tilde{\nu}_\bfz(T_v)\right]{\cal S}_{n}(T_v).
\ee
From Eq.~(\ref{TtT}),
$
T_v=T+\eta mv^2/k_B.
$
As $v$ increases, $T_v\to T_s$, $\tilde{\nu}_\bfz(T_v)\to\tilde{\nu}_\bfz(T_s)=0$, and the system ends up in the thermal equilibrium state ${\cal S}_n(T_s)$, at rest with respect to $\Lambda$. The critical velocity can be inferred from the equation $T_{v_{\rm cr}}=T_s$ which yields
\be\label{vcrit}
v_{\rm cr}(T)=\sqrt{k_B(T_s-T)/(\eta m)}.
\ee
For He II, with $T_s=2.17$ K, $v_{\rm cr}(0)=(67/\sqrt{\eta})$ m/s $\geq$ 67 m/s.

One can qualitatively understand why friction affects the moving normal fluid but not the superfluid. Friction (viscosity) can be described as a spatially  random uncorrelated perturbation. The eigenstates forming the superfluid state must be strongly entangled and, thus, resistent to such type of perturbations. Friction can only influence small separated groups of particles and excite them, at the expense of the kinetic energy of the flow, to a higher energy. Because its effect extends to the entire volume, the overall increase in energy is of order $N$. When these groups relax, they emit incoherent radiation a part of which leaves the system, another part is reabsorbed and heats it. The suitable separated small groups of particles can be found in a macroscopic number only in the eigenstates $\psi_{\Q+Nm\v/\hbar,n}=e^{i(Nm/\hbar)\v\cdot\overline{\x}}\psi_{\Q,n}$ with $|\Q|\sim\sqrt{N}$, which contribute to ${\cal S}^\v_n$.

\vspace{3pt}
\noindent
{\em Comment on Landau's critical velocity.}--We show that Landau's critical velocity~\cite{Lan,BaPe,Kad} was obtained through an erroneous argument. In Section 4 of Ref.~\cite{Lan}, Landau investigated the stability of the superfluid ground state of liquid helium in a capillary against low-energy excitations that a flow of velocity $v$ can create by losing kinetic energy. His argument is based on the properties of $\epsilon_\Q=E_{\Q,0}-E_{\bfz,0}$, the energy gap to the lowest-lying eigenstate of momentum $\hbar\Q$. In He II, $\epsilon_\Q$ is measured up to about 4\AA$^{-1}$. Its qualitative features do not depend on $T(<T_s)$: the curve starts linearly, passes over a maximum and exhibits the roton minimum~\cite{CW,Grif,Leg}. If $\epsilon_\Q$ could be measured at $T=0$, it would certainly show the same features as in the measurement~\cite{CW}, done at 1.1 K. In our notations, Landau wrote down the equation
\be\label{Lan1941}
\epsilon_{\q+Nm\v/\hbar}=\epsilon_\q+\hbar\q\cdot\v+\frac{1}{2}Nm\v^2,
\ee
first with $\epsilon_\q\approx c\hbar|\q|$ near $\q=\bfz$ [Eq. (4,1)] and second, with $\epsilon_\q\approx\Delta+(2\mu)^{-1}\hbar^2(\q-\q_r)^2$ [Eq. (4,3)] near the roton minimum $\q_r$~\cite{rem2}. He considered Eq.~(\ref{Lan1941}) as describing the energy balance of the moving fluid, so, according to him, to excite the system at the expense of a part of the kinetic energy,
\be\label{Lanineq}
\epsilon_\q+\hbar\q\cdot\v<0\quad\mbox{or}\quad v>\frac{\epsilon_\q}{\hbar|\q|}\quad\mbox{for some $\q$}
\ee
must hold. In particular, rotons are excitable if $v>\epsilon_{\q_r}/(\hbar|\q_r|)$, and phonons are excitable if $v>c$. However, the velocity threshold~(\ref{Lanineq}) has no special meaning in a flow experiment. If we add $E_{\bfz,0}$ to both sides of Eq.~(\ref{Lan1941}), we obtain
\be\label{Galn=0}
E_{\q+Nm\v/\hbar,0}=E_{\q,0}+\hbar\q\cdot\v+\frac{1}{2}Nm\v^2,
\ee
the first of Eqs.~(\ref{Galilei}) with $n=0$ and $\k=m\v/\hbar\in\Lambda^*$. By Eq.~(\ref{Galn=0}), the set $\{E_{\q,0}+\hbar\q\cdot\v+\frac{1}{2}Nm\v^2\}_{\q\in\Lambda^*}$ is the same as $\{E_{\q,0}\}_{\q\in\Lambda^*}$, with a permutation of the elements, which changes as $\v$ varies. Since the energy spectrum is unchanged, both the ground state energy and the free energy are independent of $\v$. As seen from Eq.~(\ref{Galilei+}), for any $\v$ the minimum energy $E_{\bfz,0}$ is assigned to $\psi_{Nm\v/\hbar,0}=e^{i(Nm/\hbar)\v\cdot\overline{\x}}\psi_{\bfz,0}$, the lowest-lying eigenfunction that describes the system moving with velocity $\v$, while the energy of $\psi_{\bfz,0}$ is increased by $Nm\v^2/2$. They only exchanged energies, Landau's flow is nondissipative. In flow experiments $\epsilon_\Q$ plays no role: when the normal fluid is present, the real flow is dissipative and, at any $v>0$, its loss of kinetic energy is of order $N$, highly above $\epsilon_\Q$ from $|\Q|=0$ to $|\Q|\sim\sqrt{N}$. Thus, if $\epsilon_\Q$ should be compared with the available excitation energy to get the critical velocity, then this would be zero. Nevertheless, condition~(\ref{Lanineq}) is relevant when bulk He II is locally perturbed by moving in it a tiny object (e.g. an ion) with velocity $\v$. Then, the energy transfer is indeed of order 1, and rotons and phonons are created above the respective velocity thresholds.

\vspace{3pt}
\noindent
{\em Final note.}--At $0<T<T_s$, $\epsilon_\Q$ is qualitatively the same as at $T=0$, but the energy of the state at $\Q=\bfz$ is in the order of $N$ higher than the energy of the ground state. This must be the superfluid state at the given temperature, providing an experimental support to our claim that the superfluid is a state of zero total momentum.

\vspace{3pt}
\noindent
{\em Summary.}--We studied the distribution of the total momentum of interacting quantum systems in continuous space. We derived a few mathematical results and combined them with an experimental fact, the existence of finite-momentum excitations, to make a bold intuitive jump into suggesting that the thermodynamic phases could be identified on the basis of the distribution of the total momentum in the limit of infinite volume. Different total momenta correspond to different states whose sum, weighted with their probabilities, gives the density matrix. Suitably devised measurements can project the system into states present in the density matrix. We made specific suggestions about the momentum distribution of fluids, crystals and superfluids, discussed the frictionless and the dissipative flow of the latter, derived a formula for the critical velocity, and pointed to a failure in the derivation of Landau's criterion.

\vspace{5pt}
\noindent
{\em Acknowledgement.} This work was supported by OTKA Grant No. K109577.


\begin{thebibliography}{99}
\bibitem{rem4}
If $\P\sim\sqrt{N}$, then $\p_j$ are of order 1. In a semiclassical picture, colliding particles in 1D pass through each other with unchanged momenta, if the pair potential is integrable at the origin, or exchange their momenta, if the interaction is nonintegrable. Because they are indistinguishable, it is as if they would not interact.
\bibitem{rem3}
In Ref.~\cite{Su2} the "$\g$ frame" moves with a velocity $-\hbar\g/m$ with respect to $\Lambda$. To be in accordance with the present notation, in Ref.~\cite{Su2} one should replace $\g$ by $\v$ in the upper indices, but substitute $\g=-m\v/\hbar$ in the formulas.
\bibitem{rem}
In Ref.~\cite{Su1} we show that the temperature cannot be lower in the superfluid state than in the normal one, otherwise the superfluid would immediately disappear.
\bibitem{rem2}
Landau set $\q_r=\bfz$, maybe thinking that rotons belonged to a higher branch of excitations.  From the experiments, today we know, that the roton minimum is on the "one phonon" branch at $|\q_r|\approx 2$ \AA$^{-1}$, and $\Delta\approx 10$ K, cf. Ref.~\cite{CW}.
\end{thebibliography}
\end{document}